
\hsize=13cm
\hfuzz=20pt
\vsize=18cm
\tolerance=10000
\magnification=1200

\font\headfont=cmbx10 scaled 1200
\font\namefont=cmr10
\font\initialfont=cmr10 scaled 1200
\font\addfont=cmti10

\font\fntefont= cmr7 scaled 1200
\def\fracfont#1{{\the\scriptfont0 #1}}


\def\sp{\ }
\def\seq{\sp=\sp}
\def\pls{\sp+\sp}
\def\mi{\sp-\sp}
\def\pd{\partial}
\def\mbox#1#2{\vcenter{\hrule width#1in\hbox{\vrule height#2in
   \hskip#1in\vrule height#2in}\hrule width#1in}}

\def\frac#1#2{{\fracfont{{#1}\over{#2}}}}
\def\frc#1#2{\leavevmode\kern .1em
             \raise .5ex\hbox{\the\scriptfont0 $#1$}\kern -.1em /
             \kern-.15em\lower .25ex\hbox{\the\scriptfont0 $#2$}}
\def\half{\frac{1}{2}}

\def\inv#1{\frac{1}{#1}}
\def\fder#1#2{\frac{\pd #1}{\pd #2}}
\def\der#1#2#3{\frac{{\pd}^{#1} #2}{\pd {#3}^{#1}}}
\def\mder#1#2#3{\frac{{\pd}^{#1} #2}{\pd {#3}}}
\def\oplu#1{\leavevmode\raise .3em\hbox{$\oplus$}
            \kern-1.51em\lower.4em\hbox{\the\scriptfont0 #1}}
\def\nrmord#1{\leavevmode\raise .3em\hbox{\the\scriptfont0 $#1$}
            \kern-.43em\lower.3em\hbox{\the\scriptfont0 $#1$}}

\def\vev#1{\left. #1 \right|}

\def\slash#1{\hbox{$#1$\kern-.52em\hbox{$/$}}}

%
%
\def\fnote#1#2{\baselineskip 14pt{\footnote{$^#1$}{\fntefont #2}}
               \baselineskip 18pt}
\count101=0
\def\fnoter#1{\advance\count101 by 1
              \baselineskip 14pt{\footnote{$^{\backslash \number\count101
              }$}
                                            {\fntefont #1}}
               \baselineskip 18pt}
%
%
\def\title#1#2#3{\centerline{\bf{\headfont #1}}
			\medskip
			\centerline{\bf{\headfont #2}}
			\medskip
			\centerline{\bf{\headfont #3}}
		       }
\def\contract{{\hskip -0.2truecm}
	\fnote{\star}{This work is supported in part by funds provided by the
		U. S. Department of Energy under contract\#DE-AC02-76ER03069,
 		the Swiss National Science Foundation (D.C.) and by the
		Division of Applied Mathematics of the U.S. Department of
		Energy under contract \#DE-FG02-88ER25066 (M.C.). }}
\def\author#1#2#3#4#5{\vskip 0.5truein
	\centerline{{\initialfont #1}{\namefont #2}{\initialfont #3}
		    {\initialfont#4}{\namefont#5}}
		     }
\def\authors#1#2#3#4#5{
		{{\initialfont #1}{\namefont #2}{\initialfont #3}
		    {\initialfont#4}{\namefont#5}}
			}

\def\address{\medskip
	\centerline{\addfont{Center for Theoretical Physics,}}
	\centerline{\addfont{Laboratory for Nuclear Science}}
	\centerline{\addfont{and Department of Physics,}}
	\centerline{\addfont{Massachusetts Institute of Technology}}
	\centerline{\addfont{Cambridge, Massachusetts 02139 U.S.A.}}
	    }
\def\abstract{\vskip 0.75truein
		\baselineskip 18pt plus 2pt minus 2pt
		\centerline{\bf ABSTRACT}
		\medskip
		}
\def\ctp_no#1#2{\noindent CTP \#{#1}\hfill #2}
%
%

%
%
\def\refs{\vfill\eject
	  \centerline{\bf REFERENCES}
	  \vskip 0.9truecm}
\def\ref#1#2{\item{#1.}{#2.}\smallskip\goodbreak}
\def\prd#1{{\it Phys. Rev.} {\bf D#1}}

\def\np#1{{\it Nucl. Phys.} {\bf B#1}}

\def\cqg#1{{\it Class. Quan. Grav.} {\bf #1}}

\def\prl#1{{\it Phys. Rev. Lett.} {\bf #1}}
\def\pl#1{{\it Phys. Lett.} {\bf #1B}}

\def\rmp#1{{\it Rev. Mod. Phys.} {\bf #1}}
\def\pr#1{{\it Phys. Rept.} {\bf #1}}

%
\def\a{\alpha}
\def\b{\beta}

\def\d{\delta}

\def\e{\epsilon}
\def\g{\gamma}

\def\l{\lambda}

%
%

\def\ch{{\cal H}}
\def\cj{{\cal J}}

\def\co{{\cal O}}

\baselineskip 15pt

\def\pd{\partial}

\def\ta{a}

\def\bfx{{\bf X}}

\def\dvdx{\frac{\pd V}{\pd\bfx}}
\def\edens{\frac{\e^{\a\b\g}}{\sqrt{-g}}}

\def\phihat{{\hat \phi}}
\def\Ahat{{\hat A}}
\def\Ehat{{\hat E}}


\def\RefACH{1}
\def\RefWIT{2}
\def\RefMSZ{3}
\def\RefMIS{4}
\def\RefASH{5}
\def\RefBENG{6}
\def\RefMIK{7}
\def\RefTFT{8}


\def\EqX{1}
\def\EqDREI{2}
\def\EqCOV{3}
\def\EqSG{4}
\def\EqEQM{5}
\def\EqTB{6}
\def\EqMETRIC{7}
\def\EqHAM{8}
\def\EqMINE{9}
\def\EqVREQ{10}
\def\EqDV{11}
\def\EqEGRAV{12}
\def\EqV{13}
\def\EqVAC{14}
\def\EqCURV{15}
\def\EqPEEQ{16}
\def\EqSGP{17}
\def\EqSMC{18}
\def\EqPAPP{19}
\def\EqEQMM{20}
\def\EqMSOL{21}
\def\EqEE{22}
\def\EqPLANCK{23}
\def\EqCOSMO{24}
\def\EqSC{25}
\def\EqFC{26}

\title{EINSTEIN GRAVITY IN 2+1 DIMENSIONS}
	{FROM A GAUGE MODEL WITH SYMMETRY BREAKING\contract}
	{}

\vskip 0.25truein
\centerline{
\authors{R}{OGER}{}{B}{ROOKS,}
\authors{D}{ANIEL}{}{C}{ANGEMI} and
\authors{M}{ICHAEL}{}{C}{RESCIMANNO}
	}
\address
\abstract
Einstein gravity in 2+1 dimensions arises as a consequence of
the equations of motion of a gauge model in an external metric. Newton's
constant appears as an order parameter of a
spontaneously broken discrete symmetry.
Matter is coupled in a straightforward way.
\vskip 2truecm
\centerline{Submitted to: {\it Physical Review Letters}}
\vskip 3truecm
\ctp_no{2129}{August 1992}
\vfill\eject

{\noindent {\bf Introduction.}\qquad} The Chern-Simons
$ISO(1,2)$ gauge theory (CS) approach to
gravity in 2+1 dimensions (see ref. [\RefACH,\RefWIT])
has advanced the notion that the  background metric  is
disparate from the field playing the role of the quantum
metric.
In this approach the two fields are  present in the
quantum (gauge
fixed) theory and do not arise from a background-quantum
splitting.
Additionally, the fact that the spin-connection and
dreibein appear as
independent fields also played a crucial role in the
development of the theory.
As advocated in ref. [\RefWIT], these points were key
to the
renormalizability  of the theory.  One would eventually
like to
couple the theory to matter but, thus far, attempts at
this have proven intractable. In this paper, we will be
less ambitious and simply
investigate the issue of how to introduce Newton's constant
 (or equivalently,
the Planck length) into the gauge-theoretic formulation of
$(2+1)$-dimensional
gravity.  This will lead us to a new model of
$(2+1)$-dimensional gravity .

Superficially, the non-renormalizability of General
Relativity (GR) with matter can be traced to the fact
that the coupling constant is
the Planck length, $l_{\rm P}$, and is thus dimensionful.
This is unlike QED, for
example, where the coupling constant,
$\a$, is dimensionless.  If we are to find a conventional,
perturbative
quantum field theory of gravity, we must find some dimensionless
constant to
use as a parameter.  However, we must also arrive at
Einstein's equations (with $G_{\rm N}$, Newton's constant) as
the classical equations of motion from our model.
This reminds us of the electroweak theory in which the
Fermi coupling constant is
constructed out of the parameters in the Higgs potential.
In this paper, we propose a similar framework for
$(2+1)$-dimensional gravity.
In retrospect, our study is related to earlier works
where $G_{\rm N}$ is obtained via symmetry breaking [\RefMSZ].
It is not
motivated by a popular approach to quantum gravity in which
the metric is
thought of as being a Goldstone boson [\RefMIS].

Before proceeding with the construction of our model, we
point out that we
will
be forced to abandon the CS approach to $2+1$ quantum
gravity in favor of a
formulation which has similarities with Ashtekar's
construction
[\RefASH,\RefBENG,\RefMIK].  This is due to the fact that
in order to obtain
$G_{\rm N}$ via a {\it vev}, we must couple the order
parameter to the curvature  scalar.  In the CS approach
the Einstein-Hilbert action appears in the
expansion of the action in terms of the components
(dreibein and  spin-connection) of the $ISO(1,2)$ gauge
field.  However, the Lagrangian is not gauge invariant
and multiplying it by a field which is in the singlet
of the  gauge group spoils the gauge invariance of
the action.  In contrast, the
Lagrangian in Ashtekar's formulation is gauge invariant
and so this problem does not arise.  We will ignore the
question of whether or not this
gauge invariance leads to counterterms in the theory's
perturbative
expansion and thus may endanger renormalizability.
Our focus for this paper
will be on classical physics.  With $SO(1,2)$ as the
gauge group our model is rich enough to include the
spectrum of fields in $(2+1)$-dimensional GR; it is
not necessary to use the full Poincar{\' e} group,
$ISO(1,2)$. The analysis below may be carried out
with $SO(1,2)$ replaced by $SO(3)$ with suitable
re-definitions of the metrics.
\bigskip
{\noindent{\bf The Model.}\qquad}
Our model consists of three fields, $E$, $A$ and
$\phi$.  Take $E\equiv dx^\a E_\a{}^a J_a$ a
{\hbox{$1$-form}} valued in the Lie algebra of
$SO(1,2)$ transforming in the
adjoint representation. The gauge group generators
$J_a$ satisfy $[J_a,J_b]=\e_{ab}{}^cJ_c$. Furthermore,
$A$ is a gauge field and $\phi$  a $SO(1,2)$ scalar
field  which
plays the role of the order field.  The
peculiarity of the model is that it depends, a priori,
on a background metric, $g_{\a\b}$.   Note that there
will be no  kinetic
term for $g_{\a\b}$ in our action, yet we will find a
solution which imposes
Einstein's equations on this background metric.
$E$ transforms homogeneously under the Lorentz/gauge group $SO(1,2)$ as does
the field strength $F_{\a\b}\equiv\partial_{[\a}A_{\b]}+[A_\a,A_\b]$.
Diffeomorphisms transform $E$ as a $1$-form, $F$ as a $2$-form and $\phi$ as a
scalar. We construct two gauge scalars\fnote{\dagger} {The latin indices are
raised and lowered with the Minkowski metric. The greek indices are raised and
lowered with the background (curved) metric, $g_{\a\b}$.}
$$
\bfx_{\a\b}\sp\equiv\sp E_\a{}^aE_\b{}_a \ \ ,\qquad
\bfx\sp\equiv\sp \bfx_{\a\b}g^{\a\b}\seq E_\a{}^a
E^\a{}_a \ \  .\eqno(\EqX)$$
Recall the dreibein is related to the metric by the
formula
$$g_{\a\b}\seq e_\a{}^a e_{\b a}\ \ .\eqno(\EqDREI)$$
Although the gauge and Lorentz groups
are identified,  we have two distinct
covariant derivatives
$$\eqalign{
\nabla_\a V_\b{}^{a}\seq& \pd_\a V_\b{}^{a}\mi
\Gamma_{\a\b}{}^\gamma
 V_\gamma{}^{a} \pls [A_\a,V_\b]^a\ \ ,\cr
\nabla_\a^{\rm B} V_\b{}^{a}\seq& \pd_\a V_\b{}^{a}
\mi \Gamma_{\a\b}{}^\gamma
 V_\gamma{}^{a} \pls [\omega_\a, V_{\b}]^\ta\ \ ,\cr}
 \eqno(\EqCOV)$$
written here acting on a representative field $V$.
These derivatives are both
$SO(1,2)$ and general-coordinate covariant.
$\nabla_\mu^{\rm B}$ is the covariant derivative
with respect to the background metric $g_{\alpha\beta}$
of the manifold. Thus
$\Gamma_{\alpha\beta}{}^\gamma$ is the Christoffel
symbol and  $\omega_\alpha$
is the Lorentz spin-connection (defined by
$\nabla_\a^{\rm B} e_\b{}^\ta\seq 0$).
Note that $A$ and $E$ are
independent fields while the background spin-connection
$\omega$ is constrained to be a
function of the dreibein.

Our model is defined by the classical action
$$
S^{\rm G}\seq\int d^3 x\sqrt{-g}\left\{ \phi^2
E_\a{}^{a}F_{\b\g a}
\frac{\e^{\a\b\g}}{\sqrt{-g}}
 \pls \half g^{\a\b}\nabla^{\rm B}_\a\phi
 \nabla^{\rm B}_\b\phi\mi
V(\phi,\bfx)\right\}\ \ ,\eqno(\EqSG)$$
where the mass dimensions are $\{E,A,\phi\}=\{0,1,\half\}$.
The first term looks like a Chern-Simons density
since it is topological, but it differs in two respects.
It is truly gauge invariant and $F$ is not the curvature
associated with $E$ but with the gauge field $A$.
It is modelled after the
three dimensional $BF$ (Schwarz type) topological gauge
theory [\RefTFT].
Besides
the kinetic term for $\phi$ there is also a potential
(bounded from below) in the scalars $
\phi$ and $X$ which contains the only dimensionful
constants of our model.
They are necessary to build Newton's constant $G_{\rm N}$.
Although $E_\a{}^a$ appears as an auxiliary field in
our action,
we do not integrate it out of the action.   Removing
it in this way would obscure much of the subsequent
analysis. The equations of motion we obtain by varying
$S^{\rm G}$ with respect to $A$, $E$ and $\phi$ are
\def\ft{{\tilde F}}
$$\eqalignno{
\nabla_{[\a}\phi^2E_{\b]}{}^a&\seq0\ \ ,&(\EqEQM {\rm a})\cr
\phi^2 \ft_\a{}^a&\seq  2\fder{V}{\bfx} E_\a{}^a\ \ ,
&(\EqEQM {\rm b})\cr
g^{\a\b}\nabla_\a^{\rm B}\nabla_\b^{\rm B}\phi &
\seq \mi\fder{V}{\phi}  \pls 2\phi
E_\a{}^a \ft^{\a}{}_a\ \ .&(\EqEQM {\rm c})\cr}$$
where $\ft^\a\equiv F_{\b\g} \edens$.
The metric appears in our action as an external field.
Thus the full energy-momentum tensor of the model has
no ``gravitational'' contribution.   In fact, varying
the action with respect to $g^{\a\b}$ we find the
stress-energy tensor
$$
T^{\rm B}_{\a\b}\seq
\nabla^{\rm B}_\a\phi\nabla^{\rm B}_\b\phi \mi \half
g_{\a\b}(\nabla^{\rm B} \phi)^2 \pls g_{\a\b}
V(\phi,\bfx)\mi 2\dvdx \bfx_{\a\b}\ \ .\eqno(\EqTB)$$
Upon applying the equations of motion (\EqEQM), we verify
$\nabla^{{\rm B}\a}T^{\rm B}_{\a\b}=0$.

We now study the solutions of the equations of motion
(\EqEQM) which minimize the total energy.
The latter is defined for a manifold of the form
\def\IR{\rm I\kern-.18em R}
$\IR \times\Sigma$ with a time independent metric
$$ds^2\seq dt^2 \mi \g_{ij}dx^idx^j\ \ ,\eqno(\EqMETRIC)$$
where $\g_{ij}$ is an Euclidean metric on $\Sigma$.
Let $\Pi$ be the canonical momentum conjugate to $\phi$.
The Hamiltonian density is then
$$
\ch\seq \half \Pi^2 \pls \half \g^{ij}\pd_i\phi\pd_j\phi
\pls V(\phi,\bfx)\mi 2\fder{V}{\bfx} \bfx_{00}\ \ .\eqno(\EqHAM)$$
As the first two terms are positive semi-definite, we
take $\phi$ and $\bfx$ to be the constants which minimize
the last two (potential) terms.
This yields
$$\eqalign{\d\phi:\qquad&\cr\d E^i:\qquad&\cr\d E^0:
\qquad&\cr}
\eqalign{\fder{V}{\phi}&\cr\big(\fder{V}{\bfx}&\cr
\big(\fder{V}{\bfx}&\cr}
\eqalign{\mi 2\mder{2}{V}{\phi\pd\bfx}&\bfx_{00}\seq 0\ \ ,\cr
\mi2\der{2}{V}{\bfx}& \bfx_{00}\big)E_i\seq 0\ \ ,\cr
\pls2\der{2}{V}{\bfx}& \bfx_{00}\big)E_0\seq 0\ \ .\cr}
\eqno(\EqMINE)$$

In general, given a potential, there may be many solutions.
However, they need not all be degenerate.
Two classes of potentials interest us. The first class
is composed of those with\fnote{\ddagger}{The vertical
bar signifies evaluation on a minimum
energy solution.}
$\vev{\bfx_{00}} \neq 0$ and $\vev{\bfx_{ij}} \neq 0$.
The second class has $\vev{E_\a{}^a}=0$. For the first
class, then the last two equations of (\EqMINE) imply
$$\vev{\fder{V}{\bfx}} \seq 0\ \ ,\qquad \vev{\der{2}{V}
{\bfx}}\seq 0\ \ .\eqno(\EqVREQ)$$
Using these, the equations of motion (\EqEQM) lead to
$$\vev{\fder{V}{\phi}}\seq0\qquad{\rm
and}\qquad\vev{\mder{2}{V}{\phi\pd\bfx}}\seq0\ \ , \eqno(\EqDV)$$
then if $\phi$ is non-zero
$$\vev{\nabla_{[\a}E_{\b]}}\seq 0\qquad {\rm and}\qquad
\vev{E_\a{}^a\ft^{\a}{}_a}\seq 0\ \ .\eqno(\EqEGRAV)$$
To focus ideas further, we take the potential to be
$$
V(\phi,\bfx)\seq -\half \mu^2\phi^2\pls \inv{4}\l_4
\phi^4\pls
\inv{6}\l_6\phi^6
\pls \l_X(\bfx-X_0)^4\ \ ,\eqno(\EqV)$$
where $\l_X$, $\mu^2$, $\l_4$, $\l_6$ and $X_0$ are
positive constants with mass
dimensions three, two, one, zero and zero, respectively.
For this potential, the minimal-energy solution for
$\bfx$ is $X_0$ and for $\phi$ it is $\vev{\phi}=\nu$
 where
$\nu^2=(-\l_4+\sqrt{\l_4^2 + 4\mu^2\l_6})/2\l_6$.
As is
well known, the ${\bf Z}_2$ symmetry $\phi\to -\phi$
is spontaneously broken
and as it is a discrete symmetry there is no massless
Goldstone boson.
The conditions (\EqVREQ) are satisfied by
$\vev{E_\a{}^a} = \sqrt{\frac{X_0}{3}} e_\a{}^a$
so that $\vev{\bfx}=e_\a{}^a e_{\b a} g^{\a\b}=X_0$.
Observe that this directly links the solutions for
$E_\a{}^a$ to the metric on our manifold and hence
in our action.
Such a link is a mystery in the Chern-Simons approach.
It is for this reason that the potential in $\bfx$ is
important.
Henceforth, we normalize such that $X_0=3$. The first
equation in (\EqEGRAV) is solved by $\vev{A_\a}=
\omega_\a$. We summarize our minimal-energy solutions
by the equations
$$\vev{\phi}\seq \nu\ \ ,\qquad \vev{ E_\a{}^a} \seq
e_\a{}^a\ \ ,\qquad \vev{ A_\a{}^{ab}} \seq
\omega_\a{}^{ab}\ \
.\eqno(\EqVAC)$$
With the gauge field equated to the spin-connection
we have
$$\vev{F_{\a\b}{}^{a b}}\seq \pd_{[\a}
\omega_{\b]}{}^{ab}\pls
[\omega_\a,\omega_\b]^{ab}\seq R_{\a\b}{}^{ab}\ \  ,
\eqno(\EqCURV)$$
where $R_{\a\b}{}^{ab}$ is the Riemann curvature tensor
in the background
metric $g_{\a\b}$.   The second equation in (\EqEGRAV)
(which was obtained from the equations of motion
(\EqEQM c) evaluated on $\vev{\phi}$) now reads
$$R_{\a\b}\mi\half g_{\a\b} R\seq0\ \ ,\eqno(\EqPEEQ)$$
namely, Einstein's vacuum equations. Of course,
any background metric   may appear in our action,
but only those
which solve Einstein's equations will be consistent
with our ansatz. For example flat Minkowski space
for which our solution (\EqVAC) defines a
translation invariant vacuum

The second interesting class of potentials, those
for which $\vev{E_\a{}^a}=0$ also require
$\vev{\fder{V}{\phi}}=0$. Then $\vev{\phi}=\nu$
and $\vev{\ft_\a{}^a}=0$ are minimal energy solutions.
We interpret the solution with $\vev{A_\a{}^a}=0$
as the generally covariant one.  However, here we
are unable to identify $E_\a{}^a$ solutions with
the background dreibein.

As an aside, we compare our model, $S^{\rm G}$,
with the two dominant
gauge-theoretic models appearing in the literature.
In this case, we take the coupling constants in
our potential to be such that $\phi=\nu$ dominates
the path integral.  Furthermore, we restrict the
potential to be $\bfx$ independent.  Then the
action reduces to
$$S'\seq \nu^2\int d^3 xE_\a{}^a F_{\b\g a}
\e^{\a\b\g}\ \ .\eqno(\EqSGP)$$
This is the Ashtekar formulation of
$(2+1)$-dimensional gravity
as given in ref. [\RefBENG,\RefMIK].
As shown in those references,
the constraints obtained are the same as
those of the CS approach
[\RefWIT] when the spatial metric is non-degenerate;
they satisfy the
Poincar{\' e} algebra.
\bigskip
{\noindent {\bf Matter Coupling.}\qquad}
We minimally couple matter to our model such
that we get Einstein's equations, $G_{\a\b}+
\Lambda g_{\a\b}=-8\pi G_{\rm N}T^{\rm M}_{\a\b}$.
$T^{\rm M}_{\a\b}$ is the energy-momentum
tensor one would obtain from the conventional
minimal coupling of gravity to
matter.  It should not be confused with the
stress-energy tensor which is the response
of our action to a change in the background metric,
$g_{\a\b}$.

Consider the following addition to our action
(\EqSG):
$$S^{\rm M}\seq -\half \int d^3x\sqrt{-g}
\left[\bfx^{\a\b}f(\bfx) \Upsilon_{\a\b}\pls
h(\bfx) \cj\right]\ \
.\eqno(\EqSMC)$$
$\Upsilon_{\a\b}$ (symmetric tensor) and $\cj$
depend on the matter and the metric
$g_{\a\b}$ but not on $E$, $\phi$ or $A$.
$f$ and $h$ are
dimensionless functions  of $\bfx$.
As $A$ does not couple in $S^{\rm M}$, its
equation of motion is also as before
(\EqEQM a) and is satisfied by a constant
$\phi$ and $\nabla\times E=0$.
Since we choose not to couple $\phi$
to matter, its equation of motion is as before,
namely (\EqEQM c) $$g^{\a\b}\nabla_\a\nabla_\b\phi
\seq (\mu^2\pls 2 E_\a{}^a\ft^\a{}_a)\phi \mi \l_4
\phi^3\mi \l_6 \phi^5\ \ .\eqno(\EqPAPP)$$
If $|E_\a{}^a\ft^\a{}_a|\ll\mu^2$ at each point
on the manifold (we will soon see what this means
geometrically), then  $g^{\a\b}\nabla_\a\nabla_\b
\phi \approx
-\fder{V}{\phi}$ and $\phi=\nu$ is an approximate
solution.  When we compute the variation of our
new action $S\equiv
S^{\rm G}+S^{\rm M}$, with respect to $E$ and use
$\phi=\nu$
we find that (\EqEQM b) becomes
$$
\nu^2 \ft^\a{}_a\seq   [2\fder{V}{\bfx} E^\a{}_a]
\pls
f(\bfx)
\Upsilon^{\a\b}E_{\b a}  \pls \fder{f}{\bfx}
\Upsilon^{\b\g}\bfx_{\b\g}
E^\a{}_{ a}  \pls
\fder{h}{\bfx}\cj E^\a{}_a\ \ .\eqno(\EqEQMM)$$
Now, as before, $\nabla\times E=0$ is solved by
$$E_\a{}^a\seq e_\a{}^a \qquad {\rm and}\qquad
A_\a{}^{ab}\seq \omega_\a{}^{ab}\
\ .\eqno(\EqMSOL)$$
Using these, Eq. (\EqEQMM) reads
$$R_{\a\b}\mi \half g_{\a\b} R \seq -8\pi
G_{\rm N} T^{\rm M}_{\a\b}\ \ ,\eqno(\EqEE)$$
where
$$\eqalign{
T^{\rm M}_{\a\b}\seq&f(X_0)\, \Upsilon_{\a\b}
\pls g_{\a\b}\fder{f}{\bfx}(X_0)
\Upsilon_\g{}^\g\pls
\fder{h}{\bfx}(X_0) \cj g_{\a\b}\ \ ,\cr
\nu^{2}\seq& \inv{32\pi G_{\rm N}}\ \ .\cr}
\eqno(\EqPLANCK)$$
As $\nu^2$ is now given by the inverse Planck
length, $l_{\rm P}^{-1}$, it is natural
that $\mu^2=\co(l_{\rm P}^{-2})$ (note that
$\mu=\inv{32\pi G_{\rm N}}$
for $\l_4=0$).
Our condition that $\phi=\nu$ is an approximate
solution now translates into a
condition on the scalar curvature of the
manifold: $|R|\ll l_{\rm P}^{-2}$ at each
point.  When the energy-momentum tensor of
the matter is such that
$|T^{\rm M}|=\co(l_{\rm P}^{-3})$ (or equivalently,
when we start to probe the Planck length), the
excitations of $\phi$ become important.

We now give examples.  The addition to the action
of the term
$$S^{\rm M}(\Lambda)\seq -\frac{\Lambda}
{16\pi G_{\rm N}}\int d^3x \sqrt{-g} h(\bfx)\ \
,\eqno(\EqCOSMO)$$
with $\fder{h}{\bfx}\big|=1$ yields Einstein's
equations with a cosmological constant $\Lambda$.
This gives a contribution to the potential
for $E$.
Now consider a massive scalar field, $\Phi$.
We find that a minimally coupled
action is given by (\EqSMC)  with $f(\bfx)=
h(\bfx)=\half (\bfx-5)$ so that
$$S^{\rm M}(\Phi)\seq \inv{4} \int d^3 x
\sqrt{-g}(\bfx
-5)\left[\bfx^{\a\b}\nabla_\a^{\rm B}\Phi
\nabla_\b^{\rm B}\Phi \mi m^2\Phi^2\right]\ \
.\eqno(\EqSC)$$
Evaluated on our solution (\EqVAC), the
equations of motion obtained from
$S_0+S^{\rm M}(\Phi)$ are Einstein's equations
(\EqEE) and the scalar field
equation $g^{\a\b}\nabla_\a^{\rm B}
\nabla_\b^{\rm B}\Phi=-m^2\Phi$.
Similarly, the minimally coupled action for
a massive fermion is
$$S^{\rm M}(\psi)\seq -\int d^3x \sqrt{-g}
\left\{i\half \bfx^{\a\b} \left[{\bar
\psi}\g_\a\nabla_\b^{\rm B}\psi \mi
\nabla_\a^{\rm B}{\bar\psi}\g_\b\psi\right]
\mi m{\bar\psi}\psi\right\}\ \ .\eqno(\EqFC)$$
This discussion may be generalized to other types
of matter.
\bigskip
{\noindent{\bf Comments and Conclusions.}\qquad}
In order to do quantum calculations with our action
(\EqSG), we must expand
around our vacuum configurations given by (\EqVAC)
with $e_\a{}^a=\d_\a{}^a$.  In particular,
$\phi=\nu + \phihat$, $E_\a{}^a= \d_\a{}^a + \Ehat_\a{}^a$,
$A_\a{}^{ab}= \Ahat_\a{}^{ab}$,
where the hatted fields are quantum.  Using
these expressions in the action $S=S^{\rm G}
+S^{\rm M}$, we can read off propagators and
vertices for the quantum fields in a Minkowski
background.

$\vev{E_\a{}^a}\neq0$ breaks general covariance.
Alternatively, we could have taken a potential
for which $\vev{E_\a{}^a}=0$, thereby not
breaking general covariance.  For this potential,
coupling to matter still leads to Einstein's
equations as a solutions of our model.
However, in this case, the classical solution
for $E_\a{}^a$ bears no relation to the
background dreibein.

This model does have several advantages over
the Einstein-Hilbert theory.
{}From a quantum theoretic point of view, because
the metric is treated as
a background field, the measure of the path
integral of the gauge fields
is easier to define. This is not the case with
the integration over metrics
in a path integral of the Einstein-Hilbert
action where the
metric is itself considered the fundamental
quantum field.

In conclusion, we have shown that Einstein
gravity in 2+1 dimensions
may be formulated as a gauge theory coupled
to a scalar in an external metric.
Furthermore, this model
illustrates how the Planck scale can arise
from the spontaneous
breaking of a discrete symmetry.  This precludes
the existence of additional massless particles
in the low-energy spectrum.  Matter may be
coupled to
gravity in this model in a straightforward
manner.

\refs
\ref{\RefACH}{A. Achucarro and P. Townsend,
\pl{180} (1986) 89}
\ref{\RefWIT}{E. Witten, \np{311} (1988) 46}
\ref{\RefMSZ}{S. Deser and P. van Nieuwenhuizen,
\prd{10} (1974) 401; P.
Minkowski, \pl{71} (1978) 419; L. Smolin,
\np{160} (1979) 253; A. Zee,
\prl{42} (1979) 417, \prl{44} (1980) 703; additional
references may be
found in
S. L. Adler, \rmp{54} (1982) 729}
\ref{\RefMIS}{C. W. Misner, \prd{18} (1978) 4510}
\ref{\RefASH}{For reviews see A. Ashtekar,
{\sl Non-perturbative Canonical
Quantum Gravity}, (World Scientific, Singapore, 1991);
C. Rovelli, \cqg{8}
(1991) 1613}
\ref{\RefBENG}{I. Bengtsson, \pl{220} (1989) 51}
\ref{\RefMIK}{N. Manojlovi{\' c} and A. Mikovi{\' c},
``Ashtekar Formulation
of
$2+1$ Gravity on a Torus'', preprint \#'s
Imperial/TP/91-92/7 and
QMW/PH/92/7 (April 1992)}
\ref{\RefTFT}{For a review see D. Birmingham,
M. Blau, M. Rakowski and G.
Thompson, \pr{209} (1991) 129}
\bye